\documentclass[11pt]{article}

\input{preamble/preamble.tex}
\DeclareRobustCommand{\mb}[1]{\boldsymbol{#1}}

\renewcommand{\mid}{~\vert~}

\newcommand{\mbu}{\mb{u}}
\newcommand{\mbv}{\mb{v}}

\newcommand{\mbx}{\mb{x}}

\newcommand{\mbF}{\mb{F}}

\newcommand{\mbM}{\mb{M}}

\newcommand{\mbU}{\mb{U}}
\newcommand{\mbV}{\mb{V}}

\newcommand{\mbX}{\mb{X}}

\newcommand{\mbmu}{\mb{\mu}}

\newcommand{\mbLambda}{\mb{\Lambda}}

\newcommand{\mbSigma}{\mb{\Sigma}}

\newcommand{\bbE}{\mathbb{E}}

\newcommand{\cN}{\mathcal{N}}

\newcommand{\reals}{\mathbb{R}}

\title{Statistical Neuroscience in the Single Trial Limit}
\author{Alex H. Williams and Scott W. Linderman}
\date{\today}

\begin{document}

\addtocategory{asterisk}{Williams2018,Vershynin2018,Rumayantsev2020,Rabinowitz2015,Shah2020,Williams2020,Wu2018,Udell2016,Sohn2019,Triplett2020,Chandrasekaran2018,Cowley2020,Zitnik2019}

\maketitle

\begin{abstract}
\noindent
Individual neurons often produce highly variable responses over nominally identical trials, reflecting a mixture of intrinsic ``noise'' and systematic changes in the animal's cognitive and behavioral state.
Disentangling these sources of variability is of great scientific interest in its own right, but it is also increasingly inescapable as neuroscientists aspire to study more complex and naturalistic animal behaviors.
In these settings, behavioral actions never repeat themselves exactly and may rarely do so even approximately.
Thus, new statistical methods that extract reliable features of neural activity using few, if any, repeated trials are needed.
Accurate statistical modeling in this severely trial-limited regime is challenging, but still possible if simplifying structure in neural data can be exploited.
We review recent works that have identified different forms of simplifying structure---including shared gain modulations across neural subpopulations, temporal smoothness in neural firing rates, and correlations in responses across behavioral conditions---and exploited them to reveal novel insights into the trial-by-trial operation of neural circuits.
\end{abstract}

\section*{Introduction}

Widely disseminated optical and electrophysiological recording technologies now enable many research labs to simultaneously record from hundreds, if not thousands, of neurons.
Often, a first step towards characterizing the resulting datasets is to estimate the average response of all neurons across a small set of conditions.
For example, a subject may be presented different sensory stimuli (images, odors, sounds, etc.) or trained to perform different behaviors (reaching to a target, pressing a lever, etc.), each of which constitutes a different condition. 
Each condition is then repeated many times, and the neural response is averaged over these nominally identical trials to reduce noise and variability.

Despite their obvious importance, averages represent incomplete (and potentially misleading~\cite{Golowasch2002}) summaries of neural data.
Trial-to-trial variations in neural activity reflect a variety of interesting processes, including fluctuations in attention and task engagement~\cite{Spitzer1988,Ruff2014,Rabinowitz2015,Cowley2020}, changes-of-mind during decision-making~\cite{Resulaj2009,Kiani2014,Kaufman2015,Dekleva2018,Gallivan2018}, modulations of behavioral variability to promote learning~\cite{Dhawale2019}, representations of uncertainty~\cite{Masset2020}, changes-in-strategy~\cite{Roy2020}, and modified sensory processing linked to active sensing~\cite{Verhagen2007,Fontanini2008}, locomotion~\cite{Vinck2015}, and other motor movements~\cite{Stringer2019,Musall2019}.
Some of these effects may be disentangled by developing targeted experimental designs~\cite{Spitzer1988} or by regressing against behavioral covariates~\cite{Niell2010,Musall2019,Stringer2019}.
In other cases, these effects may spontaneously emerge and subside during the course of an experiment and leave little to no behavioral signature.

While the activity of individual neurons may correlate with these single-trial phenomena, such effects may be subtle and difficult to detect.
A unique advantage to collecting \textit{simultaneous} population recordings is the possibility of pooling statistical power across many individually noisy neurons to characterize short-term fluctuations and long-term drifts in neural circuit activity.
One way to approach this goal is to estimate higher-order statistics of the neural response distribution, such as the trial-to-trial response correlations between all pairs of neurons.
Unfortunately, as we discuss below, estimating second-order response properties (i.e., covariance structure) generally requires the number of trials per condition to grow super-linearly with the number of neurons, which can quickly become infeasible.
Indeed, in modern experiments, the number of simultaneously recorded neurons often outnumbers the number of trials in each behavioral condition.
% However, we will see that there are good reasons to believe this worst-case analysis is overly pessimistic, and that it can be overcome by carefully designed statistical analyses.

At the same time, there is a trend toward studying neural circuits in more ethologically relevant settings.
This involves studying rich sensory stimuli and spontaneous, unconstrained behaviors which elicit more natural patterns of neural activity.
Recent work in visual neuroscience, for instance, has measured activity in response to very diverse sets of natural images~\cite{Cadena2019,deVries2020}, with as little as two trials per image~\cite{Stringer2019_fractal}.
This is in stark contrast with classical experiments, which presented simple stimuli (e.g. oriented gratings) repeatedly over many trials.
Similar trends are present in motor neuroscience, where motion capture algorithms have been leveraged to measure spontaneous animal behaviors~\cite{Markowitz2018,Marshall2020}.
Unconstrained motor actions repeat themselves infrequently and inexactly, resulting in few ``trials'' compared with classical behavioral tasks (e.g. cued point-to-point reaches or lever presses).

\Cref{fig:noise-corr-schematic}A summarizes these trends.
We selected a small subset of papers from the past thirty years that obtained multi-neuronal recordings, and plotted the total number of trials collected against the number of free variables one could potentially try to estimate.
There are a total of $N C$ such variables for a recording of $N$ neurons across $C$ conditions. (For now, we neglect the within-trial temporal dynamics of neural responses; including these dynamics as estimatable parameters would only exacerbate the potential for statistical error.)
The greyscale background shows the expected estimation error for second-order statistics under a worst-case scenario where trial-to-trial variations are uncorrelated across neurons and conditions (i.e., variability that is high-dimensional, in a sense that we formally define below).
We observe that the number of trials has been growing more slowly than number of parameters we wish to estimate, raising the possibility that our statistical analysis will suffer from greater estimation errors.
As these trends continue, traditional experimental designs that assume large numbers of trials over a discrete set of conditions, will become an increasingly ill-suited framework for neural data analysis---under completely unconstrained and naturalistic settings, no two experiences and actions are truly identical so, in some sense, each constitutes a unique condition with exactly one trial.

The solution is to recognize that neural and behavioral data may not be as high-dimensional as they appear.
Though we may never see the exact same pattern of neural activity or motor output twice, certain features within these data may be predictable by some (potentially nonlinear) statistical model.
If such structure can be captured by a simple (i.e. low-dimensional) model, then the ratio of trials to estimated parameters becomes larger and more favorable to accurate statistical analysis.
Such simplifying structures are present in many neural datasets. For example, trial-to-trial variability may arise from a small number of internal state variables (e.g. fluctuations in attention), neuron-by-neuron covariance might be concentrated along a small number of dimensions (i.e., approximately low-rank), and natural behavior may be composed of a small number of relatively stereotyped movements.
The success of single-trial neural data analysis is predicated on identifying and exploiting these simplifying structures in neural data, as exemplified by the recent advances we highlight below.
\section*{Statistical challenges in trial-limited regimes}

\begin{figure}
\includegraphics[width=\linewidth]{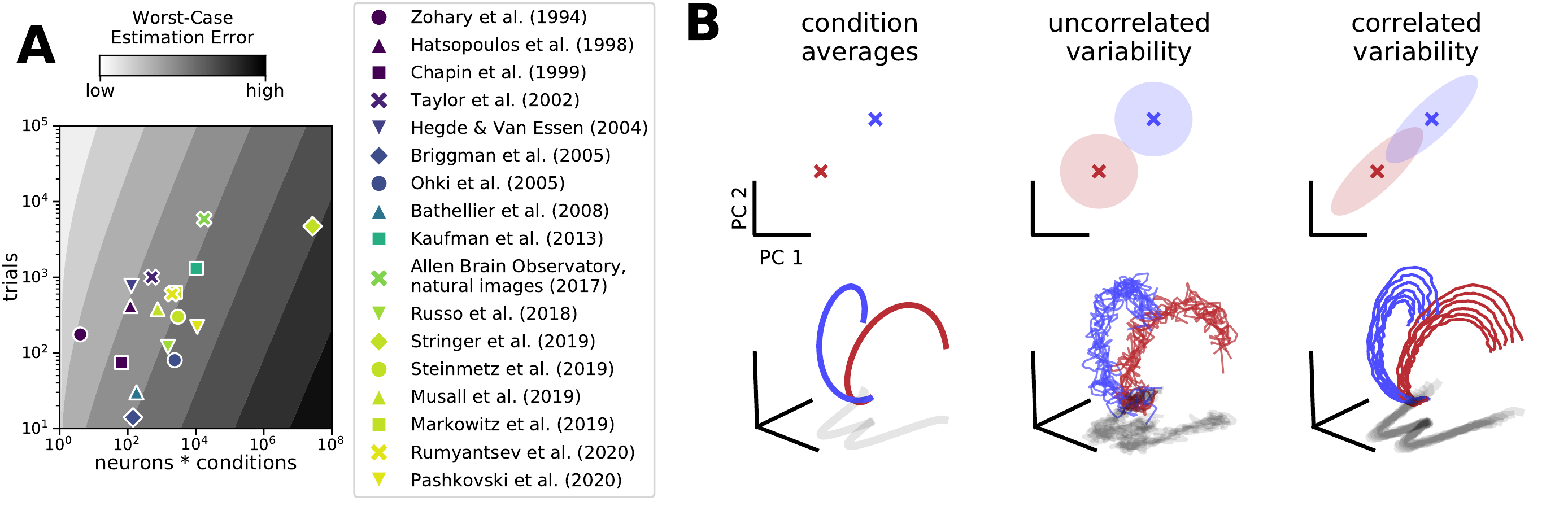}
\caption{
\textbf{(A)} The number of trials in neural datasets is growing more slowly the number of simultaneously recorded neurons and sampled behavioral conditions.
Scatter plot color corresponds to year of publication on an ordinal scale (see legend).
Grayscale heatmap shows the worst-case error scaling for covariance estimation~\cite{Vershynin2018}---the contours are $O(NC \log NC)$ for a dataset with $N$ neurons and $C$ conditions.
Darker shades correspond to larger error.
\textbf{(B)}
Low-dimensional visualizations of trial-to-trial variability in static (top row) and dynamic (bottom row) neural responses.
\textit{Left}, trial average in two conditions (blue and red). In the dynamic setting, neural firing rates evolve along a 1D curve parameterized by time. In the static setting, responses are isolated points in firing rate space.
\textit{Middle}, same responses but with independent single-trial variability illustrated in each dimension.
\textit{Right}, same responses with correlated variability.
The positive correlations in the top panel are ``information limiting'' because they increase the overlap between the two response distributions, degrading the discriminability of the two conditions (see, e.g.,~\cite{Averbeck2006}).
In the bottom panel, correlations in neural response amplitudes result in trajectories that are preferentially stretched or compressed along particular dimensions from trial-to-trial
(see~\cite{Williams2018} for a class of models that are adapted to this simplifying structure).
}
\label{fig:noise-corr-schematic}
\end{figure}

\begin{figure}
\begin{tcolorbox}[colback=red!5!white,colframe=red!75!black]
\textbf{Box 1 --- Notation}
\\[.75em]
We aimed to keep mathematical notation light, but we summarize a few points of standard notation here.
We denote scalar variables with non-boldface letters (e.g. $x$ or $X$), vectors with lowercase, boldface letters (e.g. $\mbx$), and matrices with uppercase, boldface letters (e.g., $\mbX$).
The set of real numbers is denoted by $\reals$, so the expression $s \in \reals$ means that $s$ is a scalar variable.
Likewise, $\mbv \in \reals^n$ means that $\mbv$ is a length-$n$ vector, and $\mbM \in \reals^{m \times n}$ means that $\mbM$ is a $m \times n$ matrix.
A matrix $\mbX \in \reals^{m \times n}$ is said to be ``low-rank'' if there exist matrices $\mbU \in \reals^{m \times r}$ and $\mbV \in \reals^{n \times r}$, where $r < \min(m, n)$ and for which $\mbX = \mbU \mbV^\top$.
The smallest value of $r$ for which this is possible is called the \textit{rank} of $\mbX$, in which case we would say ``$\mbX$ is a rank-$r$ matrix.''
We briefly utilize ``\textit{Big O notation}'' to represent a function up to a positive scaling constant.
More precisely, $O(g(N))$ represents any function $f(N)$ for which there exists a positive constant $C$ and a constant $N_0$ such that $|f (N)| \leq C \cdot g(N)$ for all $N \geq N_0$.
When $g$ is a monotonically increasing function of $N$, dropping constant terms like this is useful to understand the scaling behavior of the system in the limit as $N$ becomes very large.
\end{tcolorbox}
\end{figure}

In a pioneering study from the early 1990's, \textcite{Zohary1994} measured responses from co-recorded neuron pairs over 100 sessions in primates performing a visual discrimination task.
They found weak, but detectable, correlations in the neural responses over trials---when one neuron responded with a large number of spikes, the co-recorded neuron often had a slightly higher probability of emitting a large spike count.
Though this result may appear innocuous, the authors were quick to point out that even weak correlations could drastically impact the signalling capacity of sensory cortex (\cref{fig:noise-corr-schematic}B).
This finding inspired many experimental and theoretical investigations into ``noise correlations,'' which now comprise one of the most developed bodies of scientific work on trial-to-trial variability (for reviews, see~\cite{Averbeck2006,Kohn2016}).

% Though this result appears innocuous, the authors were quick to point out that these correlations could degrade the signalling capacity of the network---in essence, positive correlations among neurons with similar sensory tuning produces greater overlap in the population response, hampering the ability of an idealized decoder to recover information about the stimulus from neural activity alone (\hl{fig 2a-c}).
% This work inspired a large number of subsequent studies~\cite{Abbott1999,Kanitscheider2015}.
% Theoretical follow-ups showed, among other things, that single-trial ``noise correlations'' can also \textit{enhance} signalling capacity under appropriate conditions~\cite{Abbott1999}.

% In essence, the presence of positive correlation among neurons with similar sensory tuning produces greater overlap in the population response to different sensory stimuli, hampering the ability of an idealized decoder to recover information about the stimulus from neural activity alone.
% Although we do not illustrate it here, it is well-known that the opposite effect---i.e., that correlated variability can \textit{enhance} the fidelity of neural representations---is also possible~\cite{Abbott1999}.

The fact that \textcite{Zohary1994} were technologically limited to recording two neurons at once came with a silver lining.
For every variable of interest (i.e. a correlation coefficient between a unique pair of neurons), they collected a large number of independent trials.
%(\textasciitilde{}100 in each session, pooling across stimulus conditions).
Their statistical analyses were straightforward because the number of unknown variables was much smaller than the number of observations.
Two recent studies by \textcite{Bartolo2020} and \textcite{Rumayantsev2020} revisited this question using modern experimental techniques.
The latter group recorded calcium-gated fluorescence traces from $N \approx 1000$ neurons in mouse visual cortex over $K \approx 600$ trials.
Since each session contained roughly $N(N-1)/2 = 499500$ pairs of neurons, the number of free parameters (correlation coefficients between unique neuron pairs) was vastly larger than the number of independent observations.
Intuitively, if the correlations between neuron pair A-B and neuron pair B-C were mis-estimated, then the estimated correlation between neurons A and C would also likely be inaccurate, since the same set of trials were used for the underlying calculation.
The authors were forced to grapple with an increasingly common question: how many neurons and trials must be collected to ensure the overall conclusions were accurate?

Established results in high-dimensional statistics and random matrix theory provide answers to questions like this~\cite{Vershynin2018,Wainwright2019}.
One of these results states that $O(d \log N)$ observations (i.e., trials) are needed to accurately estimate a $N \times N$ covariance matrix, $\mbSigma$, to a specified error tolerance.
Here, $d$ measures the \textit{effective dimensionality} of the covariance matrix; formally, ${d = \mathrm{Tr}[\mbSigma] / \Vert \mbSigma \Vert}$ (see section 5.6 of~\cite{Vershynin2018}).\footnote{If the tails of the neural response distribution decay sufficiently fast, this bound can be improved from $O(d \log N)$ to $O(d)$ trials needed to accurately estimate $\mbSigma$~\cite{Vershynin2012}.}
Intuitively, $d$ is large when trial-to-trial variability equally explores every dimension of neural firing rate space (as in \cref{fig:noise-corr-schematic}B, ``uncorrelated variability'').
Conversely, $d$ is small when there are large correlations in a small number of dimensions, such that many dimensions are hardly explored relative to high-variance dimensions.
In the worst case scenario where variability is equal across all dimensions (``isotropic'') and heavy-tailed, we would have $d = N$ and require $O(N \log N)$ trials, which may be infeasible to collect.

Fortunately, \textcite{Rumayantsev2020} provide empirical evidence and a simple circuit model which suggest that the eigenvalues of $\mbSigma$ decay rapidly.
This corresponds to the statistically tractable setting where $d$ is much smaller than $N$.
\textcite{Bartolo2020} contemporaneously reported similar results under different experimental conditions and in nonhuman primates.
Overall, these works provide some of the strongest evidence to date that noise correlations do indeed limit the information content of large neural populations.
The effect of noise correlations more generally---e.g., in concert with behavioral state changes~\cite{Ni2018}, and in response to more diverse stimuli~\cite{Rikhye2015}---remains a subject of active research.

For us, the key takeaway is that the presence of simplifying structures (e.g. low dimensionality) enables accurate statistical analysis when we collect data from many more neurons and behavioral conditions than trials.
We can leverage formal results from high-dimensional statistics to sharpen this conceptual lesson into quantitative guidance for experimental designs.

\section*{Gain modulation and low-rank matrix decomposition}

\begin{figure}
\includegraphics[width=\linewidth]{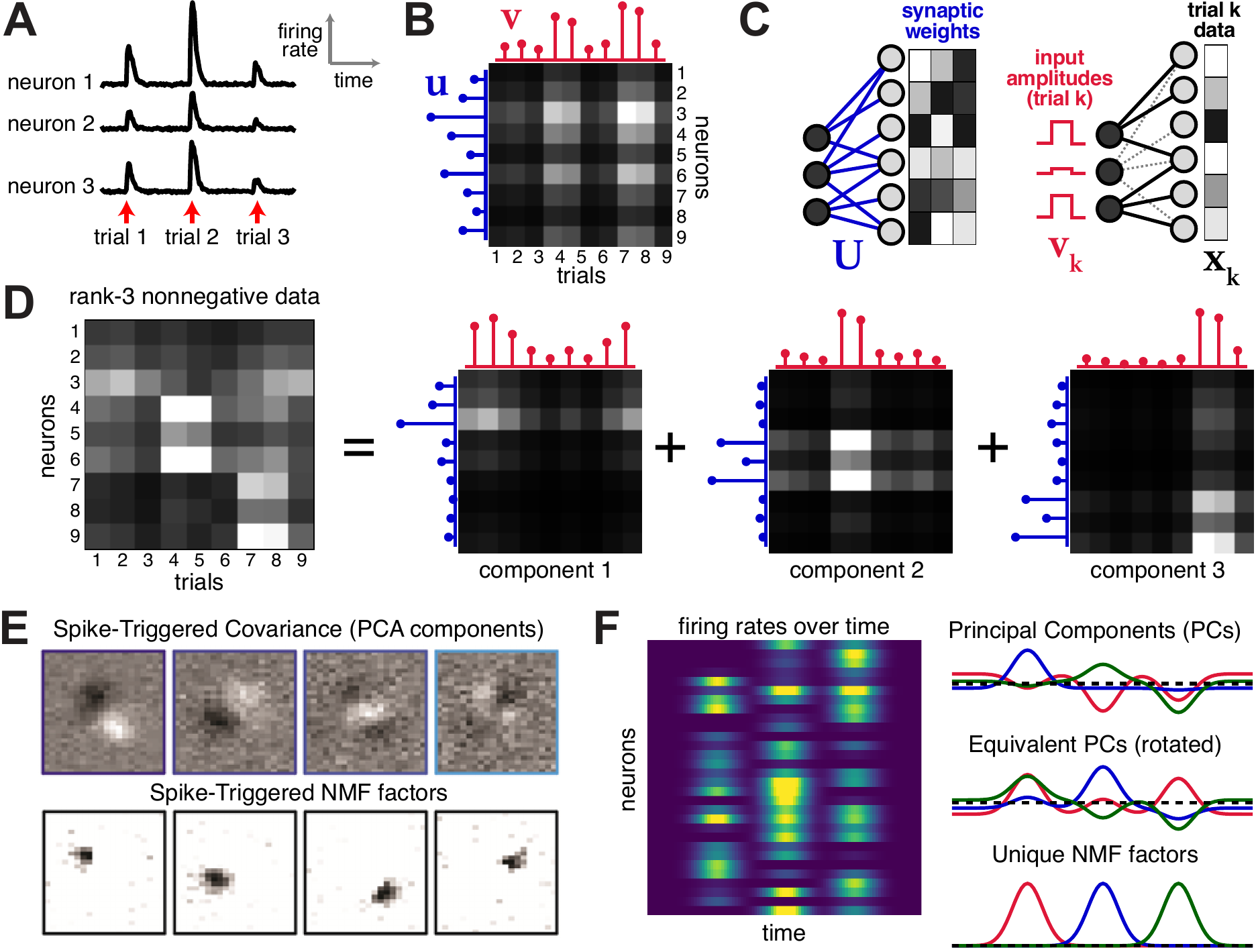}
\caption{
Matrix factorization methods for single-trial analysis.
\textbf{(A)} Schematic firing rate traces of three neurons demonstrating correlated gain-modulation: the peak responses in all three neurons are scaled by a common factor on each trial.
\textbf{(B)} A rank-1 NMF model over 9 neurons and 9 trials.
The neural responses on each trial are taken to be the peak evoked firing rate as illustrated in panel A.
The data $\mbX$ (black-to-white heatmap) are approximated by the outer product of two vectors, $\mbu \mbv^\top$ (respectively shown as blue and red stem plots).
\textbf{(C)} Interpretation of a rank-3 NMF model as an idealized neural circuit.
Low-dimensional neuron factors, $\mbU$, correspond to synaptic weights, while trial factors, $\mbv_k$ for trial $k$, correspond to input amplitudes
(a related circuit interpretation is provided in~\cite{Williams2018}).
\textbf{(D)} A schematic data matrix containing responses from 9 neurons over 9 trials is modeled as the sum of three rank-1 components (overall, a rank-3 model).
\textbf{(E)} Spike-triggered ensemble analysis of retinal ganglion neurons in Salamander retina.
NMF-identified components correspond to localized visual inputs that correspond to presynaptic bipolar cells; these signals are mixed together in PCA-identified components (panel adapted from~\cite{Liu2017}).
\textbf{(F)} Demonstration of the ``rotation problem'' in PCA.
\textit{Left}, a rank-three data matrix holding a multivariate time series.
\textit{Right}, temporal factors identified by PCA and NMF (colored lines).
Dashed black line denotes zero loading.
In this case, the decomposition by NMF is unique (up to permutations;~\cite{Donoho2004}), unlike PCA.
}
\label{fig:nmf}
\end{figure}

In the last section, we discussed covariance estimation.
It is useful to view this problem as a special case of \textit{maximum likelihood estimation} (MLE).
Given a recording of $N$ neurons, we observe neural responses $\{\mbx_1, \mbx_2, \hdots, \mbx_K\}$, where each $\mbx_k \in \reals^N$ is a vector holding the population response on trial $k \in \{1, 2, \hdots, K\}$.
Now, assume that each response is sampled independently from a multivariate normal distribution with mean $\mbmu$ and covariance $\mbSigma$; that is, $\mbx_k \sim \cN(\mbmu, \mbSigma)$ for every trial index $k$.
It is a simple exercise to show thatthe parameter estimates that maximize the log-likelihood,~$\log p (\{\mbx_1, \hdots, \mbx_K\} \mid \mbmu, \mbSigma)$, are the empirical mean,~$\widehat{\mbmu} = \frac{1}{K}\sum_{k=1}^K \mbx_k$, and the empirical covariance,~${\widehat{\mbSigma} = \frac{1}{K}\sum_{k=1}^K (\mbx_k - \widehat{\mbmu}) (\mbx_k - \widehat{\mbmu})^\top}$.

We saw that this maximum likelihood estimate of the covariance matrix is provably more accurate in trial-limited regimes when the true covariance is effectively low-dimensional \cite{Vershynin2012}.
This scenario corresponds to the covariance being approximately low-rank---i.e., there is some $N \times d$ matrix $\mbU$, for which $\mbSigma \approx \mbU \mbU^\top$.
Other modeling assumptions can achieve similar effects.
For example, \textcite{Wu2018} developed a model with \textit{Kronecker product} structure to model variability across multiple data modalities (odor conditions and neurons).
Integrating data from multiple modalities into a unified model is a nascent theme of recent research in neuroscience~\cite{Seely2016,Elsayed2017,Mishne2016,Onken2016,Williams2018}, and is already well-established in other areas of computational biology~\cite{Zitnik2019}.

Using a multivariate normal distribution to model single-trial responses is often mathematically convenient and computationally expedient.
However, it is usually not an ideal model.
For example, if we count the number of spikes in small time windows as a measure of neural activity, a Poisson distribution is typically used to model variability at the level of single neurons~\cite{Paninski2004}.
Extending the Poisson distribution to the multivariate setting turns out to be a somewhat advanced and nuanced subject~\cite{Inouye2017}.
A simple approach is to introduce per-trial \textit{latent variables}, which induce correlated fluctuations across neurons.
For example, suppose the number of spikes fired by neuron $n$ on trial $k$ is modeled as ${X_{nk} \sim \text{Poisson}(\mbu_n^\top \mbv_k)}$, where $\mbu_n \in \reals^r$ and $\mbv_k \in \reals^r$ are vectors holding $r$ ``components'' or latent variables for each neuron and trial.
It is useful to reformulate this model using matrix notation.
Let $\mbLambda = \mbU \mbV^\top$, where the matrices $\mbU \in \reals^{N \times r}$ and $\mbV \in \reals^{K \times r}$ are constructed by stacking the vectors $\mbu_n$ and $\mbv_k$, row-wise.
We can interpret $\mbLambda$ as a matrix of estimated firing rates, which, in expectation, equals the observed spikes counts---i.e., our model is that $\bbE[\mbX] = \mbLambda = \mbU \mbV^\top$, where $\mbX$ is an $N \times K$ matrix holding the observed spike counts, and noise is Poisson-distributed (i.i.d. across neurons and trials).

This model is a special case of a \textit{low-rank matrix factorization} (\textbf{Box 2})---a versatile framework that encompasses many familiar methods including PCA, k-means clustering, and others~\cite{Udell2016}.
% Roughly speaking, the model described above has the form $\mbX \approx \mbU \mbV^\top$, which is an \textit{approximate} factorization of $\mbX$ into the \textit{factor matrices} $\mbU$ and $\mbV$ (\hl{box}).
Now we ask: given $\mbX$, what are good values for $\mbU$ and $\mbV$?
One answer is to choose $\mbU$ and $\mbV$ to maximize the log-likelihood of the data.
After removing an additive constant from the log-likelihood function, we arrive at the following optimization problem:
\begin{equation}
\label{eq:nmf}
\begin{aligned}
&\text{maximize}
&&\sum_{n=1}^N \sum_{k=1}^K X_{nk} \log \Lambda_{nk} - \Lambda_{nk}
\hspace{1.5ex}
\\[1em]
&\text{subject to}
&&
\mbLambda = \mbU \mbV^\top; ~~
\mbU \geq 0; ~~
\mbV \geq 0.
\end{aligned}    
\end{equation}
We have included nonnegativity constraints which ensure that $\mbLambda \geq 0$; i.e., the predicted firing rates are, quite sensibly, nonnegative.
Alternatively, we could have dropped the nonnegativity constraints, and set $\mbLambda = \exp (\mbU \mbV^\top)$, where the exponential is applied elementwise.
However, the factor matrices are typically easier to interpret when they are nonnegative, as demonstrated by \textcite{Lee1999}, who popularized the model in \cref{eq:nmf} under the name of \textit{nonnegative matrix factorization} (NMF).
An alternative form of NMF is nearly identical, but uses a least-squares criterion instead of the Poisson log-likelihood (see~\cite{Gillis2021} for a modern review of NMF).

NMF is of immense importance to modern neural data analysis.
Variants of this method underlie a number of popular analyses: cell extraction methods in calcium imaging~\cite{Pnevmatikakis2016,Zhou2018}, sequence detection methods in neural spike trains~\cite{Peter2017,Mackevicius2019}, methods to parcellate widefield imaging videos into functional regions~\cite{Saxena2020}, and models of grid cell pattern formation~\cite{Dordek2016,Sorscher2019}.
In the context of single-trial analysis, NMF can be interpreted as a model of \textit{gain modulation} (\cref{fig:nmf}A), which is a widely studied phenomenon in sensory and motor circuits \cite{Ferguson2020,Park2020}.
For example, in a rank-1 NMF model, the firing rate matrix is factorized by a pair of vectors, $\mbLambda = \mbu \mbv^T$ (\cref{fig:nmf}B).
We can interpret $\mbu$ as being proportional to the trial-averaged firing rate of all $N$ neurons.
On trial $k$, the predicted firing rates are $\mbLambda(:, k) = v_k \mbu$, which is simply the average response re-scaled by a nonnegative gain factor, $v_k$.
It has been hypothesized that such gain modulations play a key role in tuning the signal-to-noise ratio of sensory representations, with larger gain factors corresponding to attended inputs~\cite{Reynolds2009,Rabinowitz2015}.

NMF can also model more complex patterns of single-trial variability.
In particular, an NMF model with $r$ low-dimensional components (i.e. a rank-$r$ model) can capture independent gain modulations over $r$ neural sub-populations (\cref{fig:nmf}C-D).
Despite differing in some details, recent statistical models of sensory cortex bear a strong similarity to this framework~\cite{Goris2014,Rabinowitz2015,Whiteway2017,Whiteway2019}.
We thus view this conceptual connection between NMF---a general-purpose method with many applications outside of neuroscience~\cite{Gillis2014}---and the neurobiological principle of gain modulation as a useful and unifying intuition.

Recent work has also applied NMF to \textit{spike-triggered analysis} of visually-responsive neurons.
Here, the data matrix $\mbX$ is a $S \times K$ matrix holding the spike-triggered ensemble: the $k^\text{th}$ column of $\mbX$ contains the visual stimulus (reshaped into a vector) that evoked the $k^\text{th}$ spike in the recorded neuron.
In this context, trial-to-trial variability corresponds to variation in the stimulus preceding each spike.
The classic method of spike-triggered covariance analysis~\cite{Schwartz2006}, captures this variability by a low-rank decomposition of the empirical covariance matrix---in essence, applying PCA to $\mbX$.
\textcite{Liu2017} showed that NMF extracts more interesting and physiologically interpretable structure.
When applied to data from retinal ganglion cells, NMF factors closely matched the location of presynaptic bipolar cell receptive fields, which were identified by independent experimental measurements (\cref{fig:nmf}E).
Subsequent work by \textcite{Shah2020} sharpened the spike-triggered NMF model in several respects to achieve impressive results on nonhuman primate retinal cells and V1 neurons.

The one-to-one matching of NMF factors to interpretable real-world quantities is a remarkable capability.
In contrast, the low-dimensional factors derived from PCA generally \textit{cannot} be interpreted in this manner due to the ``rotation problem'' described in \textbf{Box 2} and illustrated in \Cref{fig:nmf}F.
In essence, PCA only identifies the linear subspace containing maximal variance in the data, and there are multiple equivalent coordinate systems that describe this subspace.
Under certain conditions (outlined in~\cite{Donoho2004}), the additional constraints in the NMF objective cause the solution to be ``essentially unique'' (that is, unique up to permutations and scaling transformations).
This markedly facilitates our ability to interpret the features derived from NMF, and is one of the main explanations for the method's widespread success.

\begin{figure}
\begin{tcolorbox}[colback=red!5!white,colframe=red!75!black]
\textbf{Box 2 --- Matrix Factorization}
\\[.75em]
The expression $\widehat{\mbX} = \mbU \mbV^\top$ is called a \textit{matrix factorization} (or \textit{matrix decomposition}) of $\widehat{\mbX}$.
We call $\mbU$ and $\mbV$ \textit{factor matrices} and the columns of these matrices \textit{factors}.
This terminology is analogous to factoring natural numbers: the expression $112 = 7 \cdot 16$ is a factorization of $112$ into the factors $7$ and $16$.
Low-rank factorizations are a common element of many statistical models: given a data matrix $\mbX \in \reals^{m \times n}$, these models posit a low-rank approximation $\widehat{\mbX} = \mbU \mbV^\top$, where $\mbU \in \reals^{m \times r}$, $\mbV \in \reals^{n \times r}$ and $r < \min(m, n)$ is the rank of $\widehat{\mbX}$.
This model summarizes the $mn$ datapoints held in $\mbX$ with $mr + nr$ model parameters, which may be substantially smaller.
Intuitively, this corresponds to a form of \textit{dimensionality reduction} since the dataset is compressed into an low-dimensional (i.e., $r$-dimensional) subspace.
\\[.75em]
Adding constraints and regularization terms on the factor matrices is often very useful~\cite{Udell2016}.
For example, if $\mbF \in \reals^{r \times r}$ denotes an arbitrary invertible matrix, and if $\mbU$ and $\mbV$ are optimal factor matrices by a log-likelihood criterion, then $\mbU \mbF^{-1}$ and $\mbV \mbF^{\top}$ are also optimal factor matrices since $(\mbU \mbF^{-1}) (\mbV \mbF^\top)^\top = \mbU \mbF^{-1} \mbF \mbV^\top = \mbU \mbV^\top$.
In the context of PCA and factor analysis, this invariance is called the ``rotation problem.''
This degeneracy of solutions hinders the interpretability of the model---ideally, the pair of factor matrices that maximize the log likelihood would be unique (up to permutation and re-scaling components).
This can sometimes be accomplished by adding nonnegativity constraints (as in NMF;~\cite{Donoho2004}) or L1 regularization (as in sparse PCA;~\cite{Zou2006}).
The tensor factorization model we discuss in this review (see~\cite{Williams2018}) are also essentially unique under mild conditions.
\textcite{Kolda09} review these conditions and other forms of tensor factorizations (namely, Tucker decompositions) that do not yield unique factors.
% \\[.75em]
% For simplicity, this review largely discusses statistical modeling through the lens of maximum likelihood estimation.
% However, more sophisticated Bayesian matrix factorization models have been (see, e.g.,~\cite{Zhou2012}).
\end{tcolorbox}
\end{figure}

\section*{Single-trial variability in temporal dynamics}

\begin{figure}
\includegraphics[width=\linewidth]{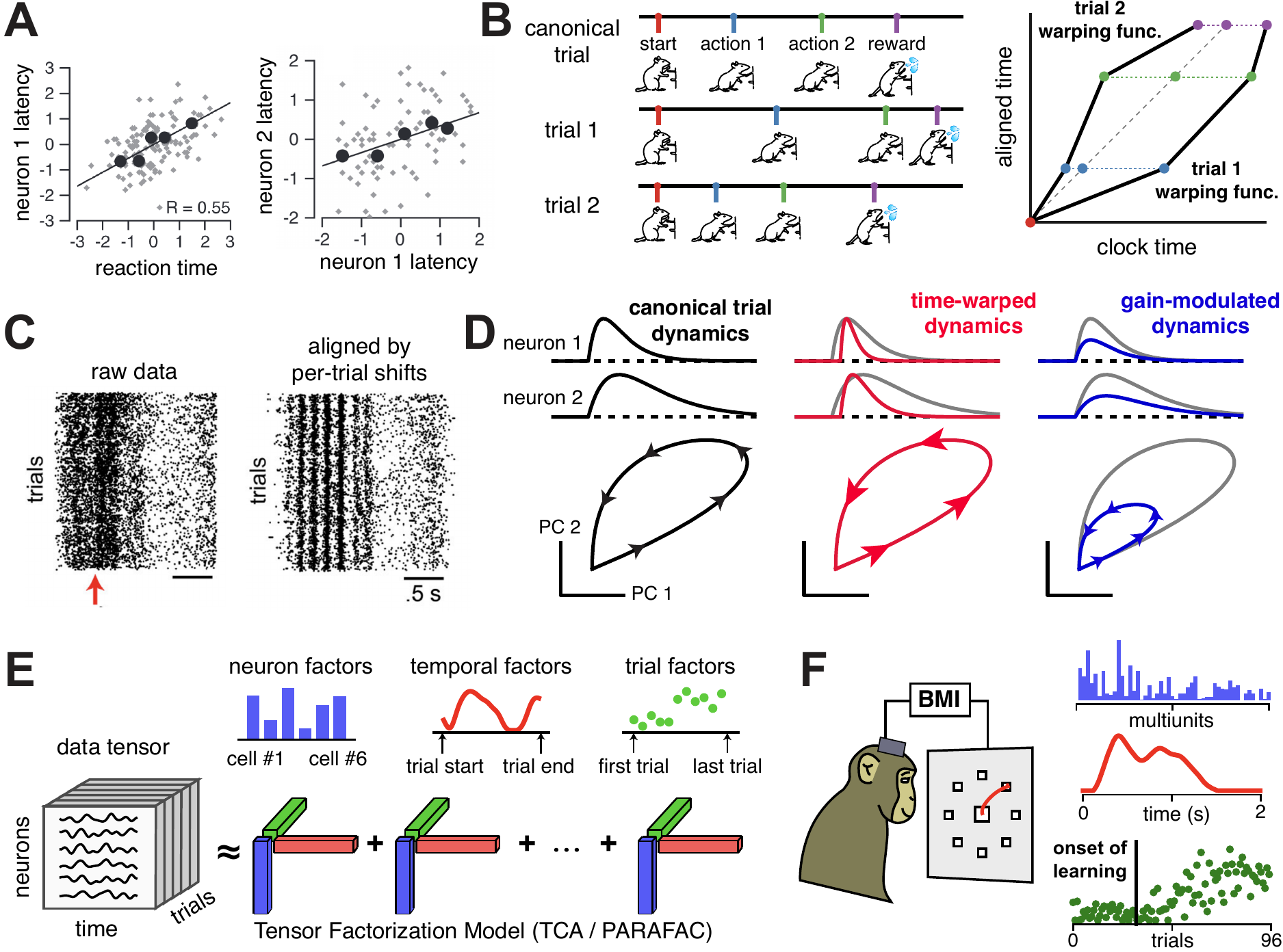}
\caption{
Models for trial-to-trial variability in temporal dynamics
\textbf{(A)} \textit{Left}, correlation between pursuit latency (reaction time) and neural response latency in a single-neuron recording from a nonhuman primate performing smooth eye pursuits. \textit{Right}, correlation in response latency between two co-recorded neurons in the same task. All units are z-scored. Large black dots denote averages over quintiles. (Adapted from~\cite{Lee2016}.)
\textbf{(B)} \textit{Left}, diagram illustrating three trials containing four behavioral actions. \textit{Right}, time warping functions that align the three trials. The ``canonical trial'' defines the identity line, and time is re-scaled on the two other trials to align each behavioral action. This manual alignment procedure is used in refs.~\cite{Leonardo2005,Kobak2016}.
\textbf{(C)} Activity from a neuron in rat motor cortex with spike times aligned to lever press (red arrow, left) and aligned by unsupervised time warping (right).
Importantly, the warping functions were fit only to neural data from simultaneously recorded neurons---the discovery of spike time oscillations demonstrates that variability in timing is correlated across cells, and thus generalizes effectively on this heldout neuron. (Adapted from~\cite{Williams2020}.)
\textbf{(D)} Illustration of how trial-to-trial variability in time warping and gain-modulated population dynamics respectively affect the speed and scale of firing rate trajectories.
Top panels show firing rate traces from two neurons, while bottom plots show the trajectory in a low-dimensional state space.
\textbf{(E)} Schematic illustration of the canonical tensor decomposition model.
\textbf{(F)} A set of three low-dimensional factors derived from tensor decomposition.
The model identifies a sub-population of neurons (blue) whose within-trial dynamics (red) grow in amplitude at the onset of learning (green).
(Adapted from~\cite{Williams2018}.)
}
\label{fig:tensor}
\end{figure}

Thus far, we have discussed models that treat neural responses as a static quantity on each trial.
However, theories spanning motor control~\cite{Vyas2020}, decision-making~\cite{Gold2007}, odor discrimination~\cite{Wilson2017}, and many other domains, all predict that the \textit{temporal} dynamics of neural circuits are crucial determinants of behavior and cognition.
Similar to how trial-to-trial variability in neural amplitudes can be described by a small number of shared gain factors, variability in temporal activity patterns also tends to be shared across neurons.
For example, the latency of neural responses can correlate with behavioral reaction times on a trial-by-trial basis.
Similarly, the latencies of co-recorded neuron pairs are often correlated~\cite{Afshar2011,Lee2016} (\cref{fig:tensor}A).

The temporal patterning of population dynamics can also vary in more complex ways.
In many experiments, each trial is composed of a sequence of sensory cues, behavioral actions, and reward dispensations.
The time delays between successive events often vary on a trial-by-trial basis, resulting in \textit{nonlinear} warping in the time course of neural dynamics (\cref{fig:tensor}B).
These misalignments can obscure salient features of the neural dynamics, and thus is it often crucial to correct for them.
This has sometimes been done in a human-supervised fashion by warping the time axis to align salient sensory and behavioral key points across trials~\cite{Leonardo2005,Kobak2016}.
Recent work has demonstrated the effectiveness of \textit{unsupervised} time warping models, which are fit purely on neural population data and are agnostic to alignment points identified by human experts~\cite{Duncker2018,Williams2020}.
Such models provide a data-driven approach for time series alignment, which can uncover unexpected features in the data---for example, \textcite{Williams2020} found \textasciitilde7 Hz oscillations in spike trains from rodent motor cortex, which were not time-locked to paw movements or the local field potential (\cref{fig:tensor}C).

% By itself, time warping only models variations in the time course of neural dynamics---if we conceptualize each trial as an $N$-dimensional trajectory in firing rate space~\cite{cite}, then time warping posits that every trial follows the same trajectory, but at different speeds (\cref{fig:tensor}D).
Beyond time warping, we are also interested in trial-to-trial changes in the trajectory's \textit{shape}.
For example, geometrical features like rotations~\cite{Churchland2012}, tangling~\cite{Russo2018}, divergence~\cite{Russo2020}, and curvature~\cite{Sohn2019}, have been used to describe population dynamics in primary and supplementary motor cortex.
While such analyses are often carried out at the level of trial averages, estimates of single-trial trajectories can be derived via well-established methods including PCA~\cite{Churchland2007}, Gaussian Process Factor Analysis (GPFA;~\cite{Yu2009}), and, more recently, artificial neural networks~\cite{Pandarinath2018}.

However, some datasets contain thousands of trials from the same population of neurons over multiple days~\cite{Dhawale2017_elife}.
In these cases, applying methods like PCA and GPFA produces thousands of estimated low-dimensional trajectories (one for each trial).
It is infeasible to visually digest and interpret such a large number of trajectories when they are overlaid on the same plot.
But if we hypothesize that trial-to-trial variation is well-described by simple transformations---e.g., time warping or gain modulation (\cref{fig:tensor}D)---we can develop statistical models that exploit this structure to reduce the dimensionality \textit{both} across trials and neurons.

\textit{Tensor factorization} (or tensor decomposition) models~\cite{Kolda09} can be used to model population activity as a small number of temporal factors that are gain-modulated across trials.
Neural data are represented in a 3-dimensional data array (called a ``tensor'') comprising neurons, timebins, and trials (\cref{fig:tensor}E, \textit{left}).
The data are then approximated by a three-way factorization, which is a straightforward generalization of the matrix factorization methods described in the last section.
This produces three intertwined sets of low-dimensional factors (\cref{fig:tensor}E, \textit{right}).
Each triplet of factors describes a sub-population of neurons (blue), with a characteristic temporal firing pattern (red), with per-trial gain modulation (green).

When successful, tensor factorization can achieve a much more aggressive degree of dimensionality reduction than matrix factorization.
For example, one demonstration in \textcite{Williams2018} on mouse prefrontal cortical dynamics shows that a tensor decomposition model utilizes 100-fold fewer parameters than a PCA model with nearly equal levels of reconstruction accuracy.
Intuitively, this dramatic reduction in dimensionality (without sacrificing performance) facilitates visualization and interpretation of the data.
For instance, \Cref{fig:tensor}F shows one low-dimensional tensor component which identifies a sub-population of neurons whose activity grew in magnitude over the course of learning in a nonhuman primate learning to adapt to a visuomotor perturbation in a brain-computer interface task.
By additionally reducing the dimensionality of the data across trials, tensor factorization is uniquely suited to pull out such trends in neural data.
Like NMF, tensor factorizations are unique under weak assumptions (see~\cite{Kolda09}), particularly if nonnegativity constraints are additionally incorporated into the model \cite{Lim2009}.
Thus, unlike PCA, tensor factorizations do not suffer from the ``rotation problem'' (see \textbf{Box 2} and \cref{fig:nmf}F), and so the extracted factors are more amenable to direct interpretation.

% Finally, the methods discussed thus far can be unified under a probabilistic framework that naturally lends itself to single trial analysis. 
Finally, low-rank and non-negative matrix factorization, time-warping models, and tensor factorizations can all be seen as probabilistic models in which each trial is endowed with latent variables that specify its unique features.
Hierarchical Bayesian models~\cite{Gelman2013-cr} generalize these notions by specifying a joint distribution over the entire dataset, combining information across trials to estimate global parameters (e.g., per-neuron factors) while simultaneously allowing latent variables (e.g., per-trial gain factors) to capture trial-by-trial variability. 
Of particular interest are hierarchical state space models~\cite{Paninski2010, Linderman2019-qj, Glaser2020}, which model each trial as a sample of a stochastic dynamical system.
For many problems of interest, we can formulate existing computational theories and hypotheses as dynamical systems models governing how a neural population's activity evolves over time \cite{Linderman2017_curr_opinion, Zoltowski2020}.
% Hierarchical state space models offer a means of capturing how those dynamics vary on a trial-by-trial basis.
% By specifying a dynamical update equation, and a noise model, these models can capture very diverse kinds of trial-to-trial variability in the temporal profile of neural firing rates, that are more complex than translations, stretching/compressing, and gain modulations.
Hierarchical state space models additionally generalize to include latent variables associated with each behavioral condition, brain region, recording session, subject, and so on.
As neuroscience progresses to study more complex and naturalistic behaviors, hierarchical models that pool statistical power while capturing variability across trials, conditions, subjects, and sessions will be a critical component of our statistical toolkit.

\section*{Open questions, challenges, and opportunities}

What scientific discoveries might statistical models with single-trial resolution help us unlock?
In general, we expect these methods to be most informative when behavioral performance is changing and unstable.
Fluctuations in attention, which are thought to correlate with population-wide gain modulations in sensory areas, represent a relatively simple and statistically tractable example that we reviewed above in detail.
More difficult tasks will often produce larger levels of trial-to-trial variability, reflecting changes in the animal's uncertainty, strategy, and appraisal of evidence, all of which may evolve stochastically over time during internal deliberation~\cite{Resulaj2009}.
Capturing signatures of these effects in neural dynamics remains challenging~\cite{Chandrasekaran2018}, and simple modeling assumptions like gain modulation and time warping may be insufficient to capture the full scope of these complexities.
State space models that optimize a set of stochastic differential equations governing neural dynamics are a promising and still evolving line of work on this subject~\cite{Zoltowski2019,Zoltowski2020}.

Trial-to-trial variability may also be heightened during incremental, long-term learning of complex tasks~\cite{Dhawale2017_review}.
This, too, represents a promising application area in which statistical methods that capture the gradual emergence of learned dynamics could connect neurobiological data to classical theories of learning.
For example, learning dynamics for hierarchically structured tasks (e.g. the categorization of objects into increasingly refined sub-categories) are expected to advance in discrete, step-like stages~\cite{Mcclelland1995}---a prediction that was recently verified in artificial deep networks~\cite{Saxe2019}.
Similar opportunities originate in theories of reinforcement learning, which have heavily influenced neuroscience research for decades~\cite{Niv2009}.
Here again, state space models offer a promising approach toward translating the constraints of competing theories into statistical models that can be tested against data~\cite{Linderman2017_curr_opinion}.

More broadly, as illustrated in \Cref{fig:noise-corr-schematic}A, we expect the field's trend towards complex experiments with trial-limited regimes to increasingly necessitate the adoption of statistical models with single-trial resolution.
Indeed, naturalistic animal behavior does not neatly sort itself into a discrete set of conditions with repeated trial structure.
We can sometimes dispense with the notion of trials altogether and characterize neural activity by directly analyzing the streaming time series of neural and behavioral data---analyses leading to the discovery of place cells, grid cells, and other functional cell types involved in navigation are prominent examples of this~\cite{Moser2008}.
However, the relationships between neural activity and behavior are not always so regular and predictable.
In many cases, it is useful to extract \textit{approximate trials} (i.e., inexact repetitions of a behavioral act or a pattern of neural population activity) from unstructured time series.
Methods such as MoSeq~\cite{Markowitz2018}, recurrent switching dynamical systems~\cite{Linderman2017}, dimensionality reduction and clustering based on wavelet features~\cite{Berman2014}, and spike sequence detection models~\cite{Williams2020_sequences,Mackevicius2019,Peter2017,Quaglio2018}, have been used for this purpose.
However, the time series clustering problem these methods aim to solve is very challenging, and it is still an active area of research.

Finally, although descriptive statistical summaries are a key step towards understanding complex neural circuit dynamics, neuroscientists should aim to integrate trial-by-trial analyses more deeply into the design and execution of causal experiments.
For example, the neural sub-populations identified by latent variable models could potentially be manipulated by emerging optogenetic stimulation protocols for large-scale populations~\cite{Marsheleaaw2019}, or targeted perturbations to brain-computer interfaces~\cite{Oby2019}.
Such interventions will be critical if we hope to build causal links between neural population dynamics and animal behavior.
Translating statistical modeling assumptions into biological terms---such as the connection between low-rank matrix factorization and gain modulation highlighted in this review---can help facilitate these interactions between experimental and theoretical research.

\subsection*{Acknowledgements}

A.H.W. received funding support from the National Institutes of Health BRAIN initiative (1F32MH122998-01), and the Wu Tsai Stanford Neurosciences Institute Interdisciplinary Scholar Program.
S.W.L. was supported by grants from the Simons Collaboration on the Global Brain (SCGB 697092) and the NIH BRAIN Initiative (U19NS113201 and R01NS113119).

% Papers in fig 1 not cited elsewhere
\nocite{Zohary1994}
\nocite{Hatsopoulos1998}
\nocite{Chapin1999}
\nocite{Taylor2002}
\nocite{Hegde2006}
\nocite{Briggman2005}
\nocite{Ohki2005}
\nocite{Bathellier2008}
\nocite{Kaufman2013}
\nocite{Russo2018}
\nocite{Stringer2019_fractal}
\nocite{Steinmetz2019}
\nocite{Musall2019}
\nocite{Markowitz2018}
\nocite{Rumayantsev2020}
\nocite{Pashkovski2020}

\printbibliography

@article{Williams2018,
  title={Unsupervised discovery of demixed, low-dimensional neural dynamics across multiple timescales through tensor component analysis},
  author={Williams, Alex H and Kim, Tony Hyun and Wang, Forea and Vyas, Saurabh and Ryu, Stephen I and Shenoy, Krishna V and Schnitzer, Mark and Kolda, Tamara G and Ganguli, Surya},
  journal={Neuron},
  volume={98},
  number={6},
  pages={1099--1115},
  year={2018},
  publisher={Elsevier},
  mynote = {\par\vspace{.25em}
    Describes tensor decomposition (``tensor components analysis'') as a general-purpose, unsupervised method for extracting within-trial and across-trial components of neural population activity.
    Diverse applications are demonstrated in data from artificial neural networks, calcium imaging data from rodent prefrontal cortex, and primate motor cortex.
    See Kolda \& Bader \cite{Kolda09} for a canonical review and introduction to tensor decompositions.
  },
}

@article{Seely2016,
    author = {Seely, Jeffrey S. AND Kaufman, Matthew T. AND Ryu, Stephen I. AND Shenoy, Krishna V. AND Cunningham, John P. AND Churchland, Mark M.},
    journal = {PLOS Computational Biology},
    publisher = {Public Library of Science},
    title = {Tensor Analysis Reveals Distinct Population Structure that Parallels the Different Computational Roles of Areas M1 and V1},
    year = {2016},
    volume = {12},
    url = {https://doi.org/10.1371/journal.pcbi.1005164},
    pages = {1-34},
    number = {11},
    doi = {10.1371/journal.pcbi.1005164}
}

@article{Mishne2016,
  title={Hierarchical coupled-geometry analysis for neuronal structure and activity pattern discovery},
  author={Mishne, Gal and Talmon, Ronen and Meir, Ron and Schiller, Jackie and Lavzin, Maria and Dubin, Uri and Coifman, Ronald R},
  journal={IEEE Journal of Selected Topics in Signal Processing},
  volume={10},
  number={7},
  pages={1238--1253},
  year={2016},
  publisher={IEEE}
}

@article{Williams2020,
  title={Discovering precise temporal patterns in large-scale neural recordings through robust and interpretable time warping},
  author={Williams, Alex H and Poole, Ben and Maheswaranathan, Niru and Dhawale, Ashesh K and Fisher, Tucker and Wilson, Christopher D and Brann, David H and Trautmann, Eric M and Ryu, Stephen and Shusterman, Roman and others},
  journal={Neuron},
  volume={105},
  number={2},
  pages={246--259},
  year={2020},
  publisher={Elsevier},
  mynote = {\par\vspace{.25em}
    Describes a flexible framework for aligning neural dynamics across trials by time warping, and demonstrates its effectiveness on several datasets spanning sensory and motor systems from rodents and primates.
    See Duncker et al. \cite{Duncker2018} for an alternative approach, and Williams \cite{Williams2020_twcpd} for notes on integrating time warping into tensor decomposition models of multi-trial data.
  },
}

@Article{Musall2019,
    author={Musall, Simon
    and Kaufman, Matthew T.
    and Juavinett, Ashley L.
    and Gluf, Steven
    and Churchland, Anne K.},
    title={Single-trial neural dynamics are dominated by richly varied movements},
    journal={Nature Neuroscience},
    year={2019},
    day={01},
    volume={22},
    number={10},
    pages={1677-1686},
    issn={1546-1726},
    doi={10.1038/s41593-019-0502-4},
    url={https://doi.org/10.1038/s41593-019-0502-4}
}

@article {Stringer2019,
	author = {Stringer, Carsen and Pachitariu, Marius and Steinmetz, Nicholas and Reddy, Charu Bai and Carandini, Matteo and Harris, Kenneth D.},
	title = {Spontaneous behaviors drive multidimensional, brainwide activity},
	volume = {364},
	number = {6437},
	elocation-id = {eaav7893},
	year = {2019},
	doi = {10.1126/science.aav7893},
	publisher = {American Association for the Advancement of Science},
	issn = {0036-8075},
	URL = {https://science.sciencemag.org/content/364/6437/eaav7893},
	eprint = {https://science.sciencemag.org/content/364/6437/eaav7893.full.pdf},
	journal = {Science}
}

@InProceedings{Linderman2017,
  title = 	 {{Bayesian Learning and Inference in Recurrent Switching Linear Dynamical Systems}},
  author = 	 {Scott Linderman and Matthew Johnson and Andrew Miller and Ryan Adams and David Blei and Liam Paninski},
  pages = 	 {914--922},
  year = 	 {2017},
  editor = 	 {Aarti Singh and Jerry Zhu},
  volume = 	 {54},
  series = 	 {Proceedings of Machine Learning Research},
  address = 	 {Fort Lauderdale, FL, USA},
  publisher =    {PMLR},
  url = 	 {http://proceedings.mlr.press/v54/linderman17a.html},
}

@article{Chandrasekaran2018,
	author = {Chandrasekaran, Chandramouli and Soldado-Magraner, Joana and Peixoto, Diogo and Newsome, William T. and Shenoy, Krishna V. and Sahani, Maneesh},
	title = {Brittleness in model selection analysis of single neuron firing rates},
	elocation-id = {430710},
	year = {2018},
	doi = {10.1101/430710},
	publisher = {Cold Spring Harbor Laboratory},
	URL = {https://www.biorxiv.org/content/early/2018/09/29/430710},
	eprint = {https://www.biorxiv.org/content/early/2018/09/29/430710.full.pdf},
	journal = {bioRxiv},
	mynote = {\par\vspace{.25em}
        The authors demonstrate that classical model selection techniques such as Akaike and Bayesian information criteria (AIC and BIC) can be surprisingly brittle and sensitive to model mismatch when analyzing single neuron recordings. To avoid these challenges, the authors argue that neuroscientists should apply and evaluate a broad variety of models, rather than a small number of pre-ordained hypotheses. Further, quantitative statistics that summarize model performance should be combined with data visualization and other qualitative measures of model agreement.
    },
}

@Article{Zoltowski2019,
  author    = {Zoltowski, David M and Latimer, Kenneth W and Yates, Jacob L and Huk, Alexander C and Pillow, Jonathan W},
  title     = {Discrete stepping and nonlinear ramping dynamics underlie spiking responses of LIP neurons during decision-making},
  journal   = {Neuron},
  year      = {2019},
  volume    = {102},
  number    = {6},
  pages     = {1249-1258},
  publisher = {Elsevier},
}

@InProceedings{Zoltowski2020,
  author    = {Zoltowski, David M and Pillow, Jonathan W and
  Linderman, Scott W},
  title     = {A general recurrent state space framework for modeling neural dynamics during decision-making},
  booktitle = {Proceedings of the 37th International Conference on Machine Learning},
  year      = {2020},
  editor    = {Hal DaumÃ© III and Aarti Singh},
  volume    = {119},
  series    = {Proceedings of Machine Learning Research},
  pages     = {11680--11691},
  address   = {Virtual},
  publisher = {PMLR},
  url       = {http://proceedings.mlr.press/v119/zoltowski20a.html},
}

@article{Mackevicius2019,
    article_type = {journal},
    title = {Unsupervised discovery of temporal sequences in high-dimensional datasets, with applications to neuroscience},
    author = {Mackevicius, Emily L and Bahle, Andrew H and Williams, Alex H and Gu, Shijie and Denisenko, Natalia I and Goldman, Mark S and Fee, Michale S},
    editor = {Colgin, Laura and Behrens, Timothy E},
    volume = 8,
    year = 2019,
    pub_date = {2019-02-05},
    pages = {e38471},
    citation = {eLife 2019;8:e38471},
    doi = {10.7554/eLife.38471},
    url = {https://doi.org/10.7554/eLife.38471},
    journal = {eLife},
    issn = {2050-084X},
    publisher = {eLife Sciences Publications, Ltd},
}

@article{Churchland2007,
  title={Techniques for extracting single-trial activity patterns from large-scale neural recordings},
  author={Churchland, Mark M and Byron, M Yu and Sahani, Maneesh and Shenoy, Krishna V},
  journal={Current opinion in neurobiology},
  volume={17},
  number={5},
  pages={609--618},
  year={2007},
  publisher={Elsevier}
}

@incollection{Duncker2018,
    title = {Temporal alignment and latent Gaussian process factor inference in population spike trains},
    author = {Duncker, Lea and Sahani, Maneesh},
    booktitle = {Advances in Neural Information Processing Systems 31},
    editor = {S. Bengio and H. Wallach and H. Larochelle and K. Grauman and N. Cesa-Bianchi and R. Garnett},
    pages = {10445--10455},
    year = {2018},
    publisher = {Curran Associates, Inc.},
    url = {http://papers.nips.cc/paper/8245-temporal-alignment-and-latent-gaussian-process-factor-inference-in-population-spike-trains.pdf}
}

@incollection{Peter2017,
title = {Sparse convolutional coding for neuronal assembly detection},
author = {Peter, Sven and Kirschbaum, Elke and Both, Martin and Campbell, Lee and Harvey, Brandon and Heins, Conor and Durstewitz, Daniel and Diego, Ferran and Hamprecht, Fred A},
booktitle = {Advances in Neural Information Processing Systems 30},
editor = {I. Guyon and U. V. Luxburg and S. Bengio and H. Wallach and R. Fergus and S. Vishwanathan and R. Garnett},
pages = {3675--3685},
year = {2017},
publisher = {Curran Associates, Inc.},
url = {http://papers.nips.cc/paper/6958-sparse-convolutional-coding-for-neuronal-assembly-detection.pdf}
}

@article{Kobak2016,
    article_type = {journal},
    title = {Demixed principal component analysis of neural population data},
    author = {Kobak, Dmitry and Brendel, Wieland and Constantinidis, Christos and Feierstein, Claudia E and Kepecs, Adam and Mainen, Zachary F and Qi, Xue-Lian and Romo, Ranulfo and Uchida, Naoshige and Machens, Christian K},
    editor = {van Rossum, Mark CW},
    volume = 5,
    year = 2016,
    pub_date = {2016-04-12},
    pages = {e10989},
    citation = {eLife 2016;5:e10989},
    doi = {10.7554/eLife.10989},
    url = {https://doi.org/10.7554/eLife.10989},
    journal = {eLife},
    issn = {2050-084X},
    publisher = {eLife Sciences Publications, Ltd},
}

@article{Rumayantsev2020,
	Author = {Rumyantsev, Oleg I. and Lecoq, J{\'e}r{\^o}me A. and Hernandez, Oscar and Zhang, Yanping and Savall, Joan and Chrapkiewicz, Rados{\l}aw and Li, Jane and Zeng, Hongkui and Ganguli, Surya and Schnitzer, Mark J.},
	Da = {2020/04/01},
	Doi = {10.1038/s41586-020-2130-2},
	Id = {Rumyantsev2020},
	Journal = {Nature},
	Number = {7801},
	Pages = {100--105},
	Title = {Fundamental bounds on the fidelity of sensory cortical coding},
	Ty = {JOUR},
	Url = {https://doi.org/10.1038/s41586-020-2130-2},
	Volume = {580},
	Year = {2020},
    mynote = {\par\vspace{.25em}
        By simultaneously recording from very large neural populations, the authors demonstrate the negative impact of correlated noise on single-trial decoding of oriented gratings in visual cortex.
        This provides direct experimental insights into a longstanding topic of interest.
        In addition, the authors provide an insightful theoretical analysis---under certain assumptions, they show that as more neurons are simultaneously recorded, \textit{fewer} trials are needed to approximate the important eigenvalues of the noise covariance.
        See the contemporaneous study by Bartolo et al. \cite{Bartolo2020}, for a similar result in a different animal model and behavioral task.
    },
}

@article{Zohary1994,
	Author = {Zohary, Ehud and Shadlen, Michael N. and Newsome, William T.},
	Da = {1994/07/01},
	Doi = {10.1038/370140a0},
	Id = {Zohary1994},
	Journal = {Nature},
	Number = {6485},
	Pages = {140--143},
	Title = {Correlated neuronal discharge rate and its implications for psychophysical performance},
	Ty = {JOUR},
	Url = {https://doi.org/10.1038/370140a0},
	Volume = {370},
	Year = {1994}
}

@article{Vyas2020,
    author = {Vyas, Saurabh and Golub, Matthew D. and Sussillo, David and Shenoy, Krishna V.},
    title = {Computation Through Neural Population Dynamics},
    journal = {Annual Review of Neuroscience},
    volume = {43},
    number = {1},
    pages = {249-275},
    year = {2020},
    doi = {10.1146/annurev-neuro-092619-094115},
}

@book{Vershynin2018,
  title={High-dimensional probability: An introduction with applications in data science},
  author={Vershynin, Roman},
  volume={47},
  year={2018},
  publisher={Cambridge university press},
  mynote={\par\vspace{.25em}
  This book packs an extensive introduction to high-dimensional statistics into fewer than 300 pages, while also providing a highly approachable and an engaging read. We believe this is a must-read for theoretical neuroscientists with an interest in the statistical foundations of single-trial analysis. We also recommend Martin Wainwright's related book \cite{Wainwright2019}, which covers many additional topics and applications.}
}

@book{Wainwright2019,
    place={Cambridge}, series={Cambridge Series in Statistical and Probabilistic Mathematics}, title={High-Dimensional Statistics: A Non-Asymptotic Viewpoint}, DOI={10.1017/9781108627771}, publisher={Cambridge University Press}, author={Wainwright, Martin J.}, year={2019}, collection={Cambridge Series in Statistical and Probabilistic Mathematics}
}

@article{Rikhye2015,
	author = {Rikhye, Rajeev V. and Sur, Mriganka},
	title = {Spatial Correlations in Natural Scenes Modulate Response Reliability in Mouse Visual Cortex},
	volume = {35},
	number = {43},
	pages = {14661--14680},
	year = {2015},
	doi = {10.1523/JNEUROSCI.1660-15.2015},
	publisher = {Society for Neuroscience},
	issn = {0270-6474},
	URL = {https://www.jneurosci.org/content/35/43/14661},
	eprint = {https://www.jneurosci.org/content/35/43/14661.full.pdf},
	journal = {Journal of Neuroscience}
}

@article{Dhawale2019,
    title = "Adaptive Regulation of Motor Variability",
    journal = "Current Biology",
    volume = "29",
    number = "21",
    pages = "3551 - 3562.e7",
    year = "2019",
    issn = "0960-9822",
    doi = "https://doi.org/10.1016/j.cub.2019.08.052",
    url = "http://www.sciencedirect.com/science/article/pii/S0960982219311029",
    author = "Ashesh K. Dhawale and Yohsuke R. Miyamoto and Maurice A. Smith and Bence P. Ölveczky",
}

@Article{Afshar2011,
    author={Afshar, Afsheen
    and Santhanam, Gopal
    and Yu, Byron M.
    and Ryu, Stephen I.
    and Sahani, Maneesh
    and Shenoy, Krishna V.},
    title={Single-Trial Neural Correlates of Arm Movement Preparation},
    journal={Neuron},
    year={2011},
    day={11},
    publisher={Elsevier},
    volume={71},
    number={3},
    pages={555-564},
    issn={0896-6273},
    doi={10.1016/j.neuron.2011.05.047},
    url={https://doi.org/10.1016/j.neuron.2011.05.047}
}

@article{Kiani2014,
title = "Dynamics of Neural Population Responses in Prefrontal Cortex Indicate Changes of Mind on Single Trials",
journal = "Current Biology",
volume = "24",
number = "13",
pages = "1542 - 1547",
year = "2014",
issn = "0960-9822",
doi = "https://doi.org/10.1016/j.cub.2014.05.049",
url = "http://www.sciencedirect.com/science/article/pii/S0960982214006149",
author = "Roozbeh Kiani and Christopher J. Cueva and John B. Reppas and William T. Newsome"
}

@article{Dekleva2018,
	Author = {Dekleva, Brian M. and Kording, Konrad P. and Miller, Lee E.},
	Da = {2018/09/03},
	Doi = {10.1038/s41467-018-05959-y},
	Id = {Dekleva2018},
	Journal = {Nature Communications},
	Number = {1},
	Pages = {3556},
	Title = {Single reach plans in dorsal premotor cortex during a two-target task},
	Ty = {JOUR},
	Url = {https://doi.org/10.1038/s41467-018-05959-y},
	Volume = {9},
	Year = {2018}
}

@article{Kaufman2015,
    article_type = {journal},
    title = {Vacillation, indecision and hesitation in moment-by-moment decoding of monkey motor cortex},
    author = {Kaufman, Matthew T and Churchland, Mark M and Ryu, Stephen I and Shenoy, Krishna V},
    editor = {Carandini, Matteo},
    volume = 4,
    year = 2015,
    pub_date = {2015-05-05},
    pages = {e04677},
    citation = {eLife 2015;4:e04677},
    doi = {10.7554/eLife.04677},
    url = {https://doi.org/10.7554/eLife.04677},
    journal = {eLife},
    issn = {2050-084X},
    publisher = {eLife Sciences Publications, Ltd},
}

@article{Gallivan2018,
	Author = {Gallivan, Jason P. and Chapman, Craig S. and Wolpert, Daniel M. and Flanagan, J. Randall},
	Da = {2018/09/01},
	Doi = {10.1038/s41583-018-0045-9},
	Journal = {Nature Reviews Neuroscience},
	Number = {9},
	Pages = {519--534},
	Title = {Decision-making in sensorimotor control},
	Ty = {JOUR},
	Url = {https://doi.org/10.1038/s41583-018-0045-9},
	Volume = {19},
	Year = {2018}
}

@article{Vershynin2012,
    author={Vershynin, Roman},
    title={How Close is the Sample Covariance Matrix to the Actual Covariance Matrix?},
    journal={Journal of Theoretical Probability},
    year={2012},
    day={01},
    volume={25},
    number={3},
    pages={655-686},
    issn={1572-9230},
    doi={10.1007/s10959-010-0338-z},
    url={https://doi.org/10.1007/s10959-010-0338-z}
}

@Article{Sohn2019,
    author={Sohn, Hansem
    and Narain, Devika
    and Meirhaeghe, Nicolas
    and Jazayeri, Mehrdad},
    title={Bayesian Computation through Cortical Latent Dynamics},
    journal={Neuron},
    year={2019},
    day={04},
    publisher={Elsevier},
    volume={103},
    number={5},
    pages={934-947.e5},
    issn={0896-6273},
    doi={10.1016/j.neuron.2019.06.012},
    url={https://doi.org/10.1016/j.neuron.2019.06.012},
    mynote = {\par\vspace{.25em}
        The authors investigate how prior beliefs and sensory evidence are integrated into neural representations in prefrontal cortex during a sensorimotor time keeping task.
        They argue that the amount of curvature in neural firing rate trajectories can affect this integration.
        Intriguingly, and relevant to this review, the authors show evidence that trial-to-trial variability is preferentially aligned with vectors tangent to the path of the trial-averaged trajectory.
        This geometric orientation of the variability is consistent with a Bayesian model---uncertainty in the response (neural variability in the output-potent dimensions) decreases as the magnitude of accumulated evidence (position along the curved neural manifold) increases.
    },
}

@article{Dhawale2017_review,
    author = {Dhawale, Ashesh K. and Smith, Maurice A. and \"{O}lveczky, Bence P.},
    title = {The Role of Variability in Motor Learning},
    journal = {Annual Review of Neuroscience},
    volume = {40},
    number = {1},
    pages = {479-498},
    year = {2017},
    doi = {10.1146/annurev-neuro-072116-031548},
}

@article {Dhawale2017_elife,
    article_type = {journal},
    title = {Automated long-term recording and analysis of neural activity in behaving animals},
    author = {Dhawale, Ashesh K and Poddar, Rajesh and Wolff, Steffen BE and Normand, Valentin A and Kopelowitz, Evi and \"{O}lveczky, Bence P},
    editor = {King, Andrew J},
    volume = 6,
    year = 2017,
    pub_date = {2017-09-08},
    pages = {e27702},
    citation = {eLife 2017;6:e27702},
    doi = {10.7554/eLife.27702},
    url = {https://doi.org/10.7554/eLife.27702},
    journal = {eLife},
    issn = {2050-084X},
    publisher = {eLife Sciences Publications, Ltd},
}

@article{Spitzer1988,
	author = {Spitzer, H and Desimone, R and Moran, J},
	title = {Increased attention enhances both behavioral and neuronal performance},
	volume = {240},
	number = {4850},
	pages = {338--340},
	year = {1988},
	doi = {10.1126/science.3353728},
	publisher = {American Association for the Advancement of Science},
	issn = {0036-8075},
	journal = {Science}
}

@article{Ruff2014,
	author = {Ruff, Douglas A. and Cohen, Marlene R.},
	title = {Global Cognitive Factors Modulate Correlated Response Variability between V4 Neurons},
	volume = {34},
	number = {49},
	pages = {16408--16416},
	year = {2014},
	doi = {10.1523/JNEUROSCI.2750-14.2014},
	publisher = {Society for Neuroscience},
	issn = {0270-6474},
	journal = {Journal of Neuroscience}
}

@article {Rabinowitz2015,
    article_type = {journal},
    title = {Attention stabilizes the shared gain of V4 populations},
    author = {Rabinowitz, Neil C and Goris, Robbe L and Cohen, Marlene and Simoncelli, Eero P},
    editor = {Carandini, Matteo},
    volume = 4,
    year = 2015,
    pub_date = {2015-11-02},
    pages = {e08998},
    citation = {eLife 2015;4:e08998},
    doi = {10.7554/eLife.08998},
    url = {https://doi.org/10.7554/eLife.08998},
    journal = {eLife},
    issn = {2050-084X},
    publisher = {eLife Sciences Publications, Ltd},
    mynote = {\par\vspace{.25em}
        The authors develop a low-dimensional factor model to capture how attentional changes modulate neural firing rates in area V4 of visual cortex.
        The model shares many features with other highlighted work---it includes a term to capture slow drift \cite{Cowley2020}, and exploits the tensor structure of the dataset \cite{Williams2018}.
        The low-dimensional factors extracted by this model improve predictive log-likelihood over models without trial-by-trial modulations.
        Further, the extracted factors are interpretable and relate concrete changes in neural processing to attentional cues, rewards, and choices.
    },
}

@article{Shah2020,
    article_type = {journal},
    title = {Inference of nonlinear receptive field subunits with spike-triggered clustering},
    author = {Shah, Nishal P and Brackbill, Nora and Rhoades, Colleen and Kling, Alexandra and Goetz, Georges and Litke, Alan M and Sher, Alexander and Simoncelli, Eero P and Chichilnisky, EJ},
    editor = {Sharpee, Tatyana O and Gold, Joshua I},
    volume = 9,
    year = 2020,
    pub_date = {2020-03-09},
    pages = {e45743},
    citation = {eLife 2020;9:e45743},
    doi = {10.7554/eLife.45743},
    url = {https://doi.org/10.7554/eLife.45743},
    journal = {eLife},
    issn = {2050-084X},
    publisher = {eLife Sciences Publications, Ltd},
    mynote = {\par\vspace{.25em}
        The authors extend the spike-triggered NMF model developed by Liu et al. \cite{Liu2017} into a fully probabilistic clustering model with additional regularization terms and spiking nonlinearities.
        Since the cluster assignments are attributed probabilistically, the model retains a ``soft clustering'' interpretation, which is similar to NMF.
        When applied to data from retinal ganglion cells, the model partitions the receptive field into functional subunits that likely reflect bipolar cell inputs.
        When the model is fit to neighboring ganglion cells, subunits are discovered in consistent locations, presumably reflecting shared pre-synaptic inputs.
        Overall, this demonstrates the ability of nonnegative factor models to identify interpretable biological structure, purely from neural activity.
    },
}

@Article{Goris2014,
    author={Goris, Robbe L. T.
    and Movshon, J. Anthony
    and Simoncelli, Eero P.},
    title={Partitioning neuronal variability},
    journal={Nature Neuroscience},
    year={2014},
    month={Jun},
    day={01},
    volume={17},
    number={6},
    pages={858-865},
    issn={1546-1726},
    doi={10.1038/nn.3711},
    url={https://doi.org/10.1038/nn.3711}
}

@Article{Verhagen2007,
    author={Verhagen, Justus V.
    and Wesson, Daniel W.
    and Netoff, Theoden I.
    and White, John A.
    and Wachowiak, Matt},
    title={Sniffing controls an adaptive filter of sensory input to the olfactory bulb},
    journal={Nature Neuroscience},
    year={2007},
    day={01},
    volume={10},
    number={5},
    pages={631-639},
    issn={1546-1726},
    doi={10.1038/nn1892},
    url={https://doi.org/10.1038/nn1892}
}

@article{Masset2020,
    title = "Behavior- and Modality-General Representation of Confidence in Orbitofrontal Cortex",
    journal = "Cell",
    volume = "182",
    number = "1",
    pages = "112 - 126.e18",
    year = "2020",
    issn = "0092-8674",
    doi = "https://doi.org/10.1016/j.cell.2020.05.022",
    url = "http://www.sciencedirect.com/science/article/pii/S0092867420306176",
    author = "Paul Masset and Torben Ott and Armin Lak and Junya Hirokawa and Adam Kepecs",
}

@article{Paninski2010,
    author={Paninski, Liam
    and Ahmadian, Yashar
    and Ferreira, Daniel Gil
    and Koyama, Shinsuke
    and Rahnama Rad, Kamiar
    and Vidne, Michael
    and Vogelstein, Joshua
    and Wu, Wei},
    title={A new look at state-space models for neural data},
    journal={Journal of Computational Neuroscience},
    year={2010},
    day={01},
    volume={29},
    number={1},
    pages={107-126},
    issn={1573-6873},
    doi={10.1007/s10827-009-0179-x},
    url={https://doi.org/10.1007/s10827-009-0179-x}
}

@Article{Niell2010,
    author={Niell, Cristopher M.
    and Stryker, Michael P.},
    title={Modulation of Visual Responses by Behavioral State in Mouse Visual Cortex},
    journal={Neuron},
    year={2010},
    day={25},
    publisher={Elsevier},
    volume={65},
    number={4},
    pages={472-479},
    issn={0896-6273},
    doi={10.1016/j.neuron.2010.01.033},
    url={https://doi.org/10.1016/j.neuron.2010.01.033}
}

@article{Paninski2004,
  title={Maximum likelihood estimation of cascade point-process neural encoding models},
  author={Paninski, Liam},
  journal={Network: Computation in Neural Systems},
  volume={15},
  number={4},
  pages={243--262},
  year={2004},
  publisher={Taylor \& Francis}
}

@inproceedings{Wu2018,
    author = {Wu, Anqi and Pashkovski, Stan and Datta, Sandeep R and Pillow, Jonathan W},
    booktitle = {Advances in Neural Information Processing Systems},
    editor = {S. Bengio and H. Wallach and H. Larochelle and K. Grauman and N. Cesa-Bianchi and R. Garnett},
    pages = {5378--5388},
    publisher = {Curran Associates, Inc.},
    title = {Learning a latent manifold of odor representations from neural responses in piriform cortex},
    volume = {31},
    year = {2018},
    mynote = {\par\vspace{.25em}
        The authors model the responses of $N=500$ neurons in piriform cortex to $P=66$ odorant stimuli using Gaussian Process latent variable models.
        To simultaneously model fluctuations across odors and neurons on each batch of trials, they decompose the overall $NP\times NP$ covariance matrix as $\mbSigma \approx \mbSigma_1 \otimes \mbSigma_2$, where $\mbSigma_1$ is an $N \times N$ covariance matrix describing correlations betweenn neurons, and $\mbSigma_2$ is a $P \times P$ covariance matrix describing correlations between odors.
        This substantially reduces the number of parameters and is an instructive example that could be repeated in many other settings for multi-modal neural data analysis.
    },
}

@Article{Elsayed2017,
    author={Elsayed, Gamaleldin F.
    and Cunningham, John P.},
    title={Structure in neural population recordings: an expected byproduct of simpler phenomena?},
    journal={Nature Neuroscience},
    year={2017},
    day={01},
    volume={20},
    number={9},
    pages={1310-1318},
    issn={1546-1726},
    doi={10.1038/nn.4617},
    url={https://doi.org/10.1038/nn.4617}
}

@article{Triplett2020,
    author = {Triplett, Marcus A. AND Pujic, Zac AND Sun, Biao AND Avitan, Lilach AND Goodhill, Geoffrey J.},
    journal = {PLOS Computational Biology},
    publisher = {Public Library of Science},
    title = {Model-based decoupling of evoked and spontaneous neural activity in calcium imaging data},
    year = {2020},
    volume = {16},
    url = {https://doi.org/10.1371/journal.pcbi.1008330},
    pages = {1-28},
    number = {11},
    doi = {10.1371/journal.pcbi.1008330},
    mynote = {\par\vspace{.25em}
        The authors combine a nonnegative factor model and a supervised model of stimulus-driven activity to respectively capture spontaneous and evoked calcium transients in large populations of neurons. The method is very effective, while also being exceedingly simple and interpretable. It is demonstrated on datasets collected from zebrafish and mouse visual systems.
    },
}

@article{Inouye2017,
    author = {Inouye, David I. and Yang, Eunho and Allen, Genevera I. and Ravikumar, Pradeep},
    title = {A review of multivariate distributions for count data derived from the Poisson distribution},
    journal = {WIREs Computational Statistics },
    volume = {9},
    number = {3},
    pages = {e1398},
    doi = {https://doi.org/10.1002/wics.1398},
    year = {2017}
}

@article{Udell2016,
    url = {http://dx.doi.org/10.1561/2200000055},
    year = {2016},
    volume = {9},
    journal = {Foundations and Trends in Machine Learning},
    title = {Generalized Low Rank Models},
    doi = {10.1561/2200000055},
    issn = {1935-8237},
    number = {1},
    pages = {1-118},
    author = {Madeleine Udell and Corinne Horn and Reza Zadeh and Stephen Boyd},
    mynote = {\par\vspace{.25em}
        The authors synthesize a large number unsupervised learning models---PCA, NMF, k-means clustering, and others---into a common matrix factorization paradigm, and use this foundation to develop a variety of model extensions. This didactic monograph is a useful entry point to neuroscientists who are interested in navigating the plethora of models that could be exploited for neural population analysis.
    },
}

@article{Lee1999,
    author={Lee, Daniel D.
    and Seung, H. Sebastian},
    title={Learning the parts of objects by non-negative matrix factorization},
    journal={Nature},
    year={1999},
    day={01},
    volume={401},
    number={6755},
    pages={788-791},
    issn={1476-4687},
    doi={10.1038/44565},
    url={https://doi.org/10.1038/44565}
}

@article{Whiteway2017,
    author = {Whiteway, Matthew R. and Butts, Daniel A.},
    title = {Revealing unobserved factors underlying cortical activity with a rectified latent variable model applied to neural population recordings},
    journal = {Journal of Neurophysiology},
    volume = {117},
    number = {3},
    pages = {919-936},
    year = {2017},
    doi = {10.1152/jn.00698.2016},
    note ={PMID: 27927786},
}

@inproceedings{Donoho2004,
    author = {Donoho, David and Stodden, Victoria},
    booktitle = {Advances in Neural Information Processing Systems},
    editor = {S. Thrun and L. Saul and B. Sch\"{o}lkopf},
    pages = {1141--1148},
    publisher = {MIT Press},
    title = {When Does Non-Negative Matrix Factorization Give a Correct Decomposition into Parts?},
    url = {https://proceedings.neurips.cc/paper/2003/file/1843e35d41ccf6e63273495ba42df3c1-Paper.pdf},
    volume = {16},
    year = {2004}
}

@article{Zhou2018,
    article_type = {journal},
    title = {Efficient and accurate extraction of in vivo calcium signals from microendoscopic video data},
    author = {Zhou, Pengcheng and Resendez, Shanna L and Rodriguez-Romaguera, Jose and Jimenez, Jessica C and Neufeld, Shay Q and Giovannucci, Andrea and Friedrich, Johannes and Pnevmatikakis, Eftychios A and Stuber, Garret D and Hen, Rene and Kheirbek, Mazen A and Sabatini, Bernardo L and Kass, Robert E and Paninski, Liam},
    editor = {Van Essen, David C},
    volume = 7,
    year = 2018,
    pub_date = {2018-02-22},
    pages = {e28728},
    citation = {eLife 2018;7:e28728},
    doi = {10.7554/eLife.28728},
    url = {https://doi.org/10.7554/eLife.28728},
    journal = {eLife},
    issn = {2050-084X},
    publisher = {eLife Sciences Publications, Ltd},
}

@article{Dordek2016,
    article_type = {journal},
    title = {Extracting grid cell characteristics from place cell inputs using non-negative principal component analysis},
    author = {Dordek, Yedidyah and Soudry, Daniel and Meir, Ron and Derdikman, Dori},
    editor = {Frank, Michael J},
    volume = 5,
    year = 2016,
    pub_date = {2016-03-08},
    pages = {e10094},
    citation = {eLife 2016;5:e10094},
    doi = {10.7554/eLife.10094},
    url = {https://doi.org/10.7554/eLife.10094},
    journal = {eLife},
    issn = {2050-084X},
    publisher = {eLife Sciences Publications, Ltd},
}

@inproceedings{Sorscher2019,
    author = {Sorscher, Ben and Mel, Gabriel and Ganguli, Surya and Ocko, Samuel},
    booktitle = {Advances in Neural Information Processing Systems},
    editor = {H. Wallach and H. Larochelle and A. Beygelzimer and F. d\textquotesingle Alch\'{e}-Buc and E. Fox and R. Garnett},
    pages = {10003--10013},
    publisher = {Curran Associates, Inc.},
    title = {A unified theory for the origin of grid cells through the lens of pattern formation},
    url = {https://proceedings.neurips.cc/paper/2019/file/6e7d5d259be7bf56ed79029c4e621f44-Paper.pdf},
    volume = {32},
    year = {2019}
}

@Article{Liu2017,
    author={Liu, Jian K.
    and Schreyer, Helene M.
    and Onken, Arno
    and Rozenblit, Fernando
    and Khani, Mohammad H.
    and Krishnamoorthy, Vidhyasankar
    and Panzeri, Stefano
    and Gollisch, Tim},
    title={Inference of neuronal functional circuitry with spike-triggered non-negative matrix factorization},
    journal={Nature Communications},
    year={2017},
    day={26},
    volume={8},
    number={1},
    pages={149},
    issn={2041-1723},
    doi={10.1038/s41467-017-00156-9},
    url={https://doi.org/10.1038/s41467-017-00156-9}
}

@article{Saxena2020,
    author = {Saxena, Shreya AND Kinsella, Ian AND Musall, Simon AND Kim, Sharon H. AND Meszaros, Jozsef AND Thibodeaux, David N. AND Kim, Carla AND Cunningham, John AND Hillman, Elizabeth M. C. AND Churchland, Anne AND Paninski, Liam},
    journal = {PLOS Computational Biology},
    publisher = {Public Library of Science},
    title = {Localized semi-nonnegative matrix factorization (LocaNMF) of widefield calcium imaging data},
    year = {2020},
    volume = {16},
    url = {https://doi.org/10.1371/journal.pcbi.1007791},
    pages = {1-28},
    number = {4},
    doi = {10.1371/journal.pcbi.1007791}
}

@article{Pnevmatikakis2016,
    title = "Simultaneous Denoising, Deconvolution, and Demixing of Calcium Imaging Data",
    journal = "Neuron",
    volume = "89",
    number = "2",
    pages = "285 - 299",
    year = "2016",
    issn = "0896-6273",
    doi = "https://doi.org/10.1016/j.neuron.2015.11.037",
    url = "http://www.sciencedirect.com/science/article/pii/S0896627315010843",
    author = "Eftychios A. Pnevmatikakis and Daniel Soudry and Yuanjun Gao and Timothy A. Machado and Josh Merel and David Pfau and Thomas Reardon and Yu Mu and Clay Lacefield and Weijian Yang and Misha Ahrens and Randy Bruno and Thomas M. Jessell and Darcy S. Peterka and Rafael Yuste and Liam Paninski",
}

@article{Schwartz2006,
  title={Spike-triggered neural characterization},
  author={Schwartz, Odelia and Pillow, Jonathan W and Rust, Nicole C and Simoncelli, Eero P},
  journal={Journal of vision},
  volume={6},
  number={4},
  pages={13--13},
  year={2006},
  publisher={The Association for Research in Vision and Ophthalmology}
}

@article{Averbeck2006,
	Author = {Averbeck, Bruno B. and Latham, Peter E. and Pouget, Alexandre},
	Da = {2006/05/01},
	Doi = {10.1038/nrn1888},
	Id = {Averbeck2006},
	Journal = {Nature Reviews Neuroscience},
	Number = {5},
	Pages = {358--366},
	Title = {Neural correlations, population coding and computation},
	Ty = {JOUR},
	Url = {https://doi.org/10.1038/nrn1888},
	Volume = {7},
	Year = {2006}
}

@article{Bartolo2020,
	author = {Bartolo, Ramon and Saunders, Richard C. and Mitz, Andrew R. and Averbeck, Bruno B.},
	title = {Information-Limiting Correlations in Large Neural Populations},
	volume = {40},
	number = {8},
	pages = {1668--1678},
	year = {2020},
	doi = {10.1523/JNEUROSCI.2072-19.2019},
	publisher = {Society for Neuroscience},
	issn = {0270-6474},
	URL = {https://www.jneurosci.org/content/40/8/1668},
	eprint = {https://www.jneurosci.org/content/40/8/1668.full.pdf},
	journal = {Journal of Neuroscience}
}

@INCOLLECTION{Gillis2014,
  title     = "The why and how of nonnegative matrix factorization",
  booktitle = "Regularization, Optimization, Kernels, and Support Vector
               Machines",
  author    = "Gillis, Nicolas",
  editor    = "Suykens, Johan A.K., and Signoretto, Marco, and Argyriou, Andreas",
  publisher = "Chapman \& Hall/CRC",
  volume    =  12,
  pages     = "257--291",
  series    = "Machine Learning \& Pattern Recognition Series",
  year      =  2014
}

@article{Zou2006,
    author = {Hui Zou and Trevor Hastie and Robert Tibshirani},
    title = {Sparse Principal Component Analysis},
    journal = {Journal of Computational and Graphical Statistics},
    volume = {15},
    number = {2},
    pages = {265-286},
    year  = {2006},
    publisher = {Taylor & Francis},
    doi = {10.1198/106186006X113430},
}

@article{Kolda09,  
    author = {Tamara G. Kolda and Brett W. Bader}, 
    title = {Tensor Decompositions and Applications}, 
    journal = {SIAM Review}, 
    volume = {51}, 
    number = {3}, 
    pages = {455--500},
    year = {2009},
    doi = {10.1137/07070111X},
}

@Article{Whiteway2019,
   Author="Whiteway, M. R.  and Socha, K.  and Bonin, V.  and Butts, D. A. ",
   Title="{{C}haracterizing the nonlinear structure of shared variability in cortical neuron populations using latent variable models}",
   Journal="Neuron Behav Data Anal Theory",
   Year="2019",
   Volume="3",
   Number="1"
}

@article{Lee2016,
    title = "Signal, Noise, and Variation in Neural and Sensory-Motor Latency",
    journal = "Neuron",
    volume = "90",
    number = "1",
    pages = "165 - 176",
    year = "2016",
    issn = "0896-6273",
    doi = "https://doi.org/10.1016/j.neuron.2016.02.012",
    url = "http://www.sciencedirect.com/science/article/pii/S0896627316001057",
    author = "Joonyeol Lee and Mati Joshua and Javier F. Medina and Stephen G. Lisberger",
}

@article {Leonardo2005,
	author = {Leonardo, Anthony and Fee, Michale S.},
	title = {Ensemble Coding of Vocal Control in Birdsong},
	volume = {25},
	number = {3},
	pages = {652--661},
	year = {2005},
	doi = {10.1523/JNEUROSCI.3036-04.2005},
	publisher = {Society for Neuroscience},
	issn = {0270-6474},
	URL = {https://www.jneurosci.org/content/25/3/652},
	eprint = {https://www.jneurosci.org/content/25/3/652.full.pdf},
	journal = {Journal of Neuroscience},
}

@article{Fontanini2008,
    author = {Fontanini, Alfredo and Katz, Donald B.},
    title = {Behavioral States, Network States, and Sensory Response Variability},
    journal = {Journal of Neurophysiology},
    volume = {100},
    number = {3},
    pages = {1160-1168},
    year = {2008},
    doi = {10.1152/jn.90592.2008},
}

@article {Roy2020,
	author = {Roy, Nicholas A. and Bak, Ji Hyun and The International Brain Laboratory, and Akrami, Athena and Brody, Carlos D. and Pillow, Jonathan W.},
	title = {Extracting the Dynamics of Behavior in Decision-Making Experiments},
	elocation-id = {2020.05.21.109678},
	year = {2020},
	doi = {10.1101/2020.05.21.109678},
	publisher = {Cold Spring Harbor Laboratory},
	journal = {bioRxiv}
}

@article{Pandarinath2018,
	Author = {Pandarinath, Chethan and O'Shea, Daniel J. and Collins, Jasmine and Jozefowicz, Rafal and Stavisky, Sergey D. and Kao, Jonathan C. and Trautmann, Eric M. and Kaufman, Matthew T. and Ryu, Stephen I. and Hochberg, Leigh R. and Henderson, Jaimie M. and Shenoy, Krishna V. and Abbott, L. F. and Sussillo, David},
	Da = {2018/10/01},
	Doi = {10.1038/s41592-018-0109-9},
	Id = {Pandarinath2018},
	Journal = {Nature Methods},
	Number = {10},
	Pages = {805--815},
	Title = {Inferring single-trial neural population dynamics using sequential auto-encoders},
	Ty = {JOUR},
	Url = {https://doi.org/10.1038/s41592-018-0109-9},
	Volume = {15},
	Year = {2018}
}

@article{Yu2009,
    author = {Yu, Byron M. and Cunningham, John P. and Santhanam, Gopal and Ryu, Stephen I. and Shenoy, Krishna V. and Sahani, Maneesh},
    title = {Gaussian-Process Factor Analysis for Low-Dimensional Single-Trial Analysis of Neural Population Activity},
    journal = {Journal of Neurophysiology},
    volume = {102},
    number = {1},
    pages = {614-635},
    year = {2009},
}

@article{Ni2018,
	author = {Ni, A. M. and Ruff, D. A. and Alberts, J. J. and Symmonds, J. and Cohen, M. R.},
	title = {Learning and attention reveal a general relationship between population activity and behavior},
	volume = {359},
	number = {6374},
	pages = {463--465},
	year = {2018},
	doi = {10.1126/science.aao0284},
	publisher = {American Association for the Advancement of Science},
	issn = {0036-8075},
	URL = {https://science.sciencemag.org/content/359/6374/463},
	eprint = {https://science.sciencemag.org/content/359/6374/463.full.pdf},
	journal = {Science}
}

@article{Hegde2006,
    author = {Hegd\'e, Jay and Van Essen, David C.},
    title = "{A Comparative Study of Shape Representation in Macaque Visual Areas V2 and V4}",
    journal = {Cerebral Cortex},
    volume = {17},
    number = {5},
    pages = {1100-1116},
    year = {2006},
    issn = {1047-3211},
    doi = {10.1093/cercor/bhl020},
    url = {https://doi.org/10.1093/cercor/bhl020},
    eprint = {https://academic.oup.com/cercor/article-pdf/17/5/1100/17297480/bhl020.pdf},
}

@article{Golowasch2002,
    author = {Golowasch, Jorge and Goldman, Mark S. and Abbott, L. F. and Marder, Eve},
    title = {Failure of Averaging in the Construction of a Conductance-Based Neuron Model},
    journal = {Journal of Neurophysiology},
    volume = {87},
    number = {2},
    pages = {1129-1131},
    year = {2002},
    doi = {10.1152/jn.00412.2001},
}

@article{Markowitz2018,
	Author = {Markowitz, Jeffrey E. and Gillis, Winthrop F. and Beron, Celia C. and Neufeld, Shay Q. and Robertson, Keiramarie and Bhagat, Neha D. and Peterson, Ralph E. and Peterson, Emalee and Hyun, Minsuk and Linderman, Scott W. and Sabatini, Bernardo L. and Datta, Sandeep Robert},
	Booktitle = {Cell},
	Date = {2018/06/28},
	Doi = {10.1016/j.cell.2018.04.019},
	Journal = {Cell},
	M3 = {doi: 10.1016/j.cell.2018.04.019},
	Number = {1},
	Pages = {44--58.e17},
	Publisher = {Elsevier},
	Title = {The Striatum Organizes 3D Behavior via Moment-to-Moment Action Selection},
	Ty = {JOUR},
	Url = {https://doi.org/10.1016/j.cell.2018.04.019},
	Volume = {174},
	Year = {2018},
	Year1 = {2018}
}

@article{Wilson2017,
	Author = {Wilson, Christopher D. and Serrano, Gabriela O. and Koulakov, Alexei A. and Rinberg, Dmitry},
	Da = {2017/11/14},
	Doi = {10.1038/s41467-017-01432-4},
	Id = {Wilson2017},
	Journal = {Nature Communications},
	Number = {1},
	Pages = {1477},
	Title = {A primacy code for odor identity},
	Ty = {JOUR},
	Url = {https://doi.org/10.1038/s41467-017-01432-4},
	Volume = {8},
	Year = {2017}
}

@article{Gold2007,
    author = {Gold, Joshua I. and Shadlen, Michael N.},
    title = {The Neural Basis of Decision Making},
    journal = {Annual Review of Neuroscience},
    volume = {30},
    number = {1},
    pages = {535-574},
    year = {2007},
    doi = {10.1146/annurev.neuro.29.051605.113038},
}

@article{Marshall2020,
	Author = {Marshall, Jesse D. and Aldarondo, Diego E. and Dunn, Timothy W. and Wang, William L. and Berman, Gordon J. and {\"O}lveczky, Bence P.},
	Booktitle = {Neuron},
	Doi = {10.1016/j.neuron.2020.11.016},
	Journal = {Neuron},
	M3 = {doi: 10.1016/j.neuron.2020.11.016},
	Publisher = {Elsevier},
	Title = {Continuous Whole-Body 3D Kinematic Recordings across the Rodent Behavioral Repertoire},
	Ty = {JOUR},
	Url = {https://doi.org/10.1016/j.neuron.2020.11.016}
}

@article {Briggman2005,
	author = {Briggman, K. L. and Abarbanel, H. D. I. and Kristan, W. B.},
	title = {Optical Imaging of Neuronal Populations During Decision-Making},
	volume = {307},
	number = {5711},
	pages = {896--901},
	year = {2005},
	doi = {10.1126/science.1103736},
	publisher = {American Association for the Advancement of Science},
	issn = {0036-8075},
	URL = {https://science.sciencemag.org/content/307/5711/896},
	eprint = {https://science.sciencemag.org/content/307/5711/896.full.pdf},
	journal = {Science}
}

@article{Russo2020,
	Author = {Russo, Abigail A. and Khajeh, Ramin and Bittner, Sean R. and Perkins, Sean M. and Cunningham, John P. and Abbott, L. F. and Churchland, Mark M.},
	Booktitle = {Neuron},
	Date = {2020/08/19},
	Doi = {10.1016/j.neuron.2020.05.020},
	Journal = {Neuron},
	M3 = {doi: 10.1016/j.neuron.2020.05.020},
	Number = {4},
	Pages = {745--758.e6},
	Publisher = {Elsevier},
	Title = {Neural Trajectories in the Supplementary Motor Area and Motor Cortex Exhibit Distinct Geometries, Compatible with Different Classes of Computation},
	Ty = {JOUR},
	Url = {https://doi.org/10.1016/j.neuron.2020.05.020},
	Volume = {107},
	Year = {2020},
	Year1 = {2020}
}

@article{Russo2018,
    title = "Motor Cortex Embeds Muscle-like Commands in an Untangled Population Response",
    journal = "Neuron",
    volume = "97",
    number = "4",
    pages = "953 - 966.e8",
    year = "2018",
    issn = "0896-6273",
    doi = "https://doi.org/10.1016/j.neuron.2018.01.004",
    url = "http://www.sciencedirect.com/science/article/pii/S0896627318300072",
    author = "Abigail A. Russo and Sean R. Bittner and Sean M. Perkins and Jeffrey S. Seely and Brian M. London and Antonio H. Lara and Andrew Miri and Najja J. Marshall and Adam Kohn and Thomas M. Jessell and Laurence F. Abbott and John P. Cunningham and Mark M. Churchland",
}

@article{Churchland2012,
	Author = {Churchland, Mark M. and Cunningham, John P. and Kaufman, Matthew T. and Foster, Justin D. and Nuyujukian, Paul and Ryu, Stephen I. and Shenoy, Krishna V.},
	Da = {2012/07/01},
	Doi = {10.1038/nature11129},
	Id = {Churchland2012},
	Journal = {Nature},
	Number = {7405},
	Pages = {51--56},
	Title = {Neural population dynamics during reaching},
	Ty = {JOUR},
	Url = {https://doi.org/10.1038/nature11129},
	Volume = {487},
	Year = {2012}
}

@Inbook{Mcclelland1995,
    author={McClelland, J. L.},
    title={A connectionist perspective on knowledge and development.},
    series={Developing cognitive competence:  New approaches to process modeling.},
    year={1995},
    publisher={Lawrence Erlbaum Associates, Inc},
    address={Hillsdale,  NJ,  US},
    pages={157-204},
}

@article{Vinck2015,
	Author = {Vinck, Martin and Batista-Brito, Renata and Knoblich, Ulf and Cardin, Jessica A.},
	Booktitle = {Neuron},
	Date = {2015/05/06},
	Doi = {10.1016/j.neuron.2015.03.028},
	Journal = {Neuron},
	M3 = {doi: 10.1016/j.neuron.2015.03.028},
	Number = {3},
	Pages = {740--754},
	Publisher = {Elsevier},
	Title = {Arousal and Locomotion Make Distinct Contributions to Cortical Activity Patterns and Visual Encoding},
	Ty = {JOUR},
	Url = {https://doi.org/10.1016/j.neuron.2015.03.028},
	Volume = {86},
	Year = {2015},
	Year1 = {2015}
}

@article{Oby2019,
	author = {Oby, Emily R. and Golub, Matthew D. and Hennig, Jay A. and Degenhart, Alan D. and Tyler-Kabara, Elizabeth C. and Yu, Byron M. and Chase, Steven M. and Batista, Aaron P.},
	title = {New neural activity patterns emerge with long-term learning},
	volume = {116},
	number = {30},
	pages = {15210--15215},
	year = {2019},
	doi = {10.1073/pnas.1820296116},
	publisher = {National Academy of Sciences},
	issn = {0027-8424},
	URL = {https://www.pnas.org/content/116/30/15210},
	eprint = {https://www.pnas.org/content/116/30/15210.full.pdf},
	journal = {Proceedings of the National Academy of Sciences}
}

\end{document}